\documentclass[usenatbib,onecolumn]{mn2e}

\usepackage{graphicx}
\usepackage{amssymb}

\usepackage[authoryear]{natbib}
\usepackage{subfigure}
\usepackage{color}
\newcommand{\beq}{\begin{equation}}
\newcommand{\eeq}{\end{equation}}

\def\ssr{Sp Sci. Rev.}

\def\bar{\overline}

\def\mnras{MNRAS}
\def\apj{ApJ}

\def\aap{A\&A}

\def\emf{\overline{\mbox{${\cal E}$}} {}}
\def\emfb{\overline{\mbox{\boldmath ${\cal E}$}} {}}
\def\bbE{\bar {\bf E}}

\def\beq{\begin{equation}}
\def\ee{\end{equation}}
\def\lsim{\mathrel{\rlap{\lower4pt\hbox{\hskip1pt$\sim$}}
    \raise1pt\hbox{$<$}}}
\def\gsim{\mathrel{\rlap{\lower4pt\hbox{\hskip1pt$\sim$}}
    \raise1pt\hbox{$>$}}}
\def\bfE{{\bf E}}
\def\bfJ{{\bf J}}

\def\bfA{{\bf A}}
\def\bfa{{\bf a}}
\def\bfe{{\bf e}}
\def\bfB{{\bf B}}
\def\bbJ{\bar {\bf J}}
\def\bbV{\bar {\bf V}}

\def\bB{\overline B}
\def\ts{\times}
\def\lb{\langle}
\def\rb{\rangle}
\def\curl{\nabla {\ts}}

\def\bbV{\bar {\bf V}}
\def\bfv{{\bf v}}

\def\bfV{{\bf V}}
\def\bfj{{\bf j}}

\def\bfe{{\bf e}}
\def\bfw{{\bomega}}

\def\bfb{{\bf b}}

\def\bfB{{\bf B}}
\def\bfA{{\bf A}}

\def\bbB{\overline {\bf B}}
\def\bbA{\overline {\bf A}}

\def\div{\nabla\cdot}

\title[Magnetic fields as ribbons not lines: a  dynamo lesson]
 {Ribbons    characterize  magnetohydrodynamic magnetic fields better than lines:  a  lesson from dynamo theory}
\author [Eric G. Blackman and Alexander Hubbard]{Eric G. Blackman$^{1}$\thanks{E-mail: blackman@pas.rochester.edu} and Alexander Hubbard $^{2}$\thanks{E-mail:ahubbard@amnh.org }\\ $^{1}$Department of Physics and Astronomy, University of Rochester, Rochester NY, 14618, USA\\ $^{2}$Department of Astrophysics, American Museum of Natural History, New York, NY 10024-5192, USA}
\begin{document}
\date{}
\pagerange{\pageref{firstpage}--\pageref{lastpage}} \pubyear{}
\maketitle
\label{firstpage}
\begin{abstract}
Blackman \& Brandenburg 
 argued  that magnetic helicity  conservation in dynamo theory can in principle be captured by  diagrams of mean field dynamos when the magnetic fields are represented by ribbons or tubes, but not by  lines.  Here we  present  such a   schematic ribbon diagram for the  $\alpha^2$ dynamo that  tracks magnetic helicity and provides  distinct scales of large scale magnetic helicity, small scale magnetic helicity, and kinetic helicity  involved in the process.  This also  motivates  our  construction of a new ``2.5 scale''  minimalist generalization of the helicity-evolving equations for  the $\alpha^2$ dynamo that   separately allows for these three distinct length scales while keeping only two dynamical equations.
We solve these   equations and, as in previous  studies, find that  the large scale field first grows at a rate independent of the magnetic Reynolds number $R_M$ before quenching  to an $R_M$ dependent regime.  But we also show  that  the larger the ratio of the wavenumber where the  small scale current helicity resides to that of the forcing scale, the earlier the non-linear dynamo quenching occurs, and the weaker the large scale field is at the turnoff from linear growth.
 The  harmony between the theory and the schematic diagram exemplifies a general lesson that  magnetic fields in MHD are better visualized as  two-dimensional ribbons (or pairs of lines)  rather than single lines.
\end{abstract}
\section{Introduction}

Dynamo theory has long been a  topic of active research in astrophysical  and geophysical magnetohydrodynamics, and describes  how magnetic energy can be  amplified and/or sustained 
in the presence of a turbulent diffusion that would otherwise rapidly dissipate the field (e.g. Glatzmeier 2002, Brandenburg \& Subramanian 2005a, Blackman 2014).
In particular, large scale dynamo (LSD) theory  in astrophysics is aimed at  understanding the \emph{in situ} physics of magnetic field  growth, saturation, and  sustenance  on time or spatial scales large compared to the turbulent scales of the host rotator.  Galaxies, stars,  planets, compact objects, and accretion engines (via their jets)  often show direct or indirect evidence for large scale magnetic fields.
%\citep{2004astro.ph.11739S} 
The extent to which magnetic fields are a fossil vestige of the formation of the object, or primarily generated via LSDs, is itself a question  of  active research; but observations of the  sun for example (e.g. Wang \& Sheely 1993;  Schrijver  \&  Zwaan 2000) prove that LSDs do operate in nature since
%in nature since a 
a frozen-in field could not exhibit sign reversals.
%The sun exhibits sign reversals  of the large scale field over each half  solar cycle .   If the field were simply that frozen into to the ISM during initial formation, there would  be no reversals. 
%LSDs are  also thought to be   operating in galaxies, despite whatever initial fields may have been seeded %cosmologically because of the in situ processing that occurs 
%from  supernova driven turbulence during the galaxies' lifetime. 
%LSDs  
LSDs are also commonly seen in accretion disc  or shearing box simulations (e.g. Brandenburg et al. 1995; Lesur \& Ogilive 2008; Davis et al. 2009; Gressel et al. 2010; Simon et al. 2011;  Sorathia et al. 2012; Suzuki \& Inutsuka 2013; Ebrahimi \& Bhattacharjee 2014), with  observed   large scale field reversals occurring on time scales of order  $10$ orbits. 

 The presence of magnetohydrodynamic  turbulence makes the study of LSDs highly nonlinear, exacerbating  the importance of numerical simulations.  However, large scale spatial symmetry and the slow evolution of  large scale fields  compared to turbulent fluctuation  time scales motivates  a mean field approach in which statistical, spatial, or temporal averages are taken, and the evolution of the mean field studied (Moffatt 1978 Parker 1979; Krause \& R\"adler 1980). 
 For many decades, mean field dynamo theory was studied assuming a prescribed flow that was not affected by the growing magnetic field. This is  a problem because the field does exert Lorentz forces back on the flow. Moreover, unless the back reaction is  incorporated into the theory, it is impossible
 to make a dynamical prediction  of LSD  saturation.   The 21st century has brought progress in this endeavor as the  growth and saturation of LSDs seen in  simple closed box simulations (Brandenburg 2001)  are reasonably matched by newer mean field dynamo theories  that follow the time dependent dynamical evolution of magnetic helicity (e.g. Blackman \& Field 2002, for  reviews see Brandenburg \& Subramanian 2005a; Blackman 2014).  
These newer 21st century mean field dynamos  are  discrete scale theories
that capture the basic principles of the inverse transfer of the  magnetic helicity first derived in the  spectral  model of Pouquet et al. (1976), which revealed the fundamental importance
of magnetic helicity evolution for large scale field growth. 

The simplest   example of a mean field dynamo for which this saturation theory has been tested dynamically (Field \& Blackman 2002; Blackman \& Field 2002)  and compared to simulations (Brandenburg 2001)  is the $\alpha^2$ dynamo in a closed or periodic box.  In this dynamo, the system is forced with kinetic helicity in a closed or periodic system at some wavenumber, say $k_f=5$
%AH: add
%EB3: ok
where the box wavenumber is $k_1=1$.  The large scale field growth at $k_1$ is captured by an equation which is coupled to the equation for small scale magnetic helicity  because the sum of small and large scale magnetic helicity is conserved up to resistive terms.  The growth of large scale helicity thus also implies the growth of small scale helicity of the opposite sign.  Because the growth driver turns out to be the difference between
small scale kinetic and small scale current  helicities (the latter related to the small scale magnetic helicity), the growth of the small scale magnetic helicity offsets the kinetic helicity driver,  eventually halting the growth.  
While $\alpha^2$ dynamos are simple theoretical models for study, they are also  important for the generation of large scale fields in stars that lack strong differential rotation. 

Blackman and Brandenburg (2003) argued that  textbook  diagrams of mean field dynamos do not properly capture the near conservation of magnetic helicity  because
they display a growth of large scale magnetic helicity without any  corresponding oppositely signed small scale helicity to compensate.   If instead the field is displayed as  a ribbon or tube, any  writhing
%AH: do we really want the "twisting" above?
%EB3: deleted. good point.
of the large scale field is naturally accompanied by a corresponding twisting of the small scale field.
On a related theme, 
%{\bf 
Pfister \& Gekelman (1991) showed that magnetic helicity conservation during a magnetic  reconnection  of linked loops  can be captured when the field lines are represented as ribbons but not as lines. See also  Bellan (2000) in this context.
%}

In this paper we revisit the $\alpha^2$ dynamo to  show  more specifically how 
to  visually  represent its large scale and small scale magnetic helicity growth.
The resulting diagram has also stimulated us to introduce a minimalist  generalization of the helicity evolving $\alpha^2$ dynamo equations which allows for  scale separation between the large scale, forcing scale, and scale of the  small scale current helicity, whilst keeping only two dynamical equations. Normally the latter two latter scales are taken to be the same, but there is  evidence from simulations that they can be different (Park \& Blackman 2012).

In section 2 we discuss the different conceptual representations of magnetic helicity. In section 3 we discuss  the schematic diagram of the $\alpha^2$ dynamo. In  section 4 we derive the  the equations for the $\alpha^2$ dynamo based on previous work but with the new scale separation feature. We solve these equations in section  5 and discuss their solutions and the correspondence with the schematic diagram.  We conclude in section 6.

\section{Magnetic Helicity as a measure of linkage, twist,  or writhe}

Magnetic helicity is  defined as the volume integral of the
dot product of the vector potential $\bf A$ and the magnetic field ${\bf B}=\curl {\bf A}$, i.e.
$
\int \bfA\cdot \bfB  dV.
%\label{1}
$
 This is a measure of magnetic linkage (e.g. Moffatt 1978;  Berger \& Field 1984):
Consider two linked flux tubes (or ribbons whose flux is contained within an imaginary tube) of
%EB2 added the parenthesis above
cross sectional area vectors $d \bf S_1$ and $d\bf S_2$ respectively. Then $\Phi_1= \int \bfB_1\cdot d \bf S_1$ is the magnetic flux in tube 1, where $\bfB_1$ is its magnetic field.  Similarly,  the flux of the second tube is  $\Phi_2= \int \bfB_2\cdot d \bf S_2$ where $\bfB_2$ is its magnetic field. If the fields in each tube are of  constant  magnitude and parallel to $d \bf S_1$ and $d\bf S_2$ respectively, we can 
 write the magnetic helicity as the sum of  contributions from each of the two flux tube volumes to obtain
\beq
\int \bfA\cdot \bfB  dV= \int\int \bfA_1\cdot \bfB_1 dl_1 dS_1 +  \int \int\bfA_2 \cdot \bfB_2 dl_2 dS_2,
\label{link1}
\eeq
where we have factored the volume integrals into  products of line and surface integrals, with the line integrals taken along the direction parallel to $\bfB_1$ and $\bfB_2$. Since the magnitudes of $\bfB_1$ and $\bfB_2$ are constant in the tubes
%, and the tubes are axially symmetric, 
 we  can  pull $\bfB_1$ and $\bfB_2$ out of each of the two line integrals on the right of  Eq. (\ref{link1}) to write
\beq
\int \bfA\cdot \bfB  dV= \int A_1 dl_1 \int B_1dS_1 +  \int A_2  dl_2 \int B_2dS_2
= \Phi_2 \Phi_1 + \Phi_1 \Phi_2= 2\Phi_1\Phi_2,
%\int \int \bfA_1 \cdot \bfB dl_1 dS_1= \int A_1 dl_1 \int B_1 dS_1 
\label{link}
\eeq
%  $\int \int \bfA_1 \cdot \bfB dl_1 dS_1= \int A_1 dl_1 \int B_1 dS_1 = \Phi_2 \Phi_1$,
where we used Gauss' theorem to replace $\int A_1 dl_1=\Phi_2$, the magnetic flux of tube 2 that is linked through tube 1,
and similarly $\int A_2 dl_1=\Phi_1$.  If the tubes are not linked, then the line integral would vanish
and there would be no magnetic helicity.
  
Helicity can also be  characterized  as a measure of magnetic  ``twist''  and  ``writhe''.
While a more extensive discussion  of Figs. 1 and 2 will come later, note that the schematic in the  top left of Fig. 2 shows a single closed magnetic ribbon with zero net helicity. In evolving the loop from the top left to the top right, 
four right-handed writhes  have been introduced. Each of these  writhes can be perceived as  a loop through which a rigid pole can be threaded along the direction of the electromotive force (as shown in the second panels of Fig 1 and 2)  and around which the ribbons winds  in the right handed sense. However,
%EB3 just tweaked to indicate there are now two poles in both figures
no dissipation was involved in the evolution and and the overall ribbon remans closed.
Under these conditions, magnetic helicity is conserved in MHD, so one left-handed twist along each of the  writhed loops appears  as a consequence of magnetic helicity conservation:  One unit of writhe has the same amount of magnetic helicity as one full twist of the ribbon.

Having shown that linkage of flux is a measure of helicity, and argued that writhe and twist are equivalent, it remains
to establish the correspondence between linkage and either writhe or twist.
Seeing the equivalence between linkage and twist  is wonderfully aided by the use of a strip of paper to represent a magnetic  flux ribbon, along with  scissors and tape (e.g. Bellan 2000, Blackman 2014).
Give a long  strip of paper a full $360^{\circ}$ right handed twist around its long axis (by twisting clockwise at the top with your right hand while holding the bottom with your left hand)   and fasten the ends so that it is now a twisted closed loop.  Now consider that this  ribbon could  have been composed of two adjacent ribbons pressed together side by side. Separating these  
adjacent ribbons is achieved using scissors. Cut the along the center line of the strip all the way around. The result is  two linked ribbons, each of 1/2 the width of the original, and each with one right handed twist.   The amount of   helicity of the system remains conserved but has
been transformed into a different form:
if the original unseparated ribbon had a magnetic flux $\Phi$ then each of the half  ribbons  has  flux $\Phi/2$. From the above discussion of linkage,   the linkage of these two new ribbons gives a magnetic helicity $2 \Phi^2/4 =   \Phi^2/2$.   But the original uncut ribbon had a single right handed twist with total flux $\Phi$.  For the helicity of the  initial twisted ribbon to equal that of the  two linked twisted ribbons, any  twisted  ribbon must contribute a helicity equal to its flux squared.
Helicity is thus conserved:   $\Phi^2$  is the helicity associated with the initial twisted ribbon. This  equals the sum of helicity from  the linkage of the two half-thickness ribbons $\Phi^2/2$ plus that from the 
right handed twists in each of these two ribbons $2 \times (\Phi/2)^2 =  \Phi^2/2$.

%Having established the quantitative relation between twist and linkage, now consider the relation between twist and writhe:Start again with a straight paper ribbon and give the  ribbon one right handed twist around its long axis. Laying the twisted ribbon on a flat surface, push the ends  toward the other end  and the tube will buckle. Then imagine attaching the ends by sliding and stretching one end around to the other without lifting or twisting the ribbon. The  one unit twist along the ribbon has now been converted into a writhe of the ribbon establishing that one unit of twist helicity is equivalent to one unit of writhe helicity.

To summarize:  one unit of twist helicity for a ribbon or tube of magnetic flux $\Phi$  is equal to one unit of writhe helicity for the same  ribbon, and both are  separately equal to 1/2 of  the helicity resulting from  the  linkage of two untwisted flux tubes of flux $\Phi$.
These three different, but equivalent  ways of thinking about magnetic helicity are instructive for 
extracting the physical ideas in what follows.

\section{Diagrammatic Description of an $\alpha^2$ Dynamo}

Traditional 20th century  textbook dynamos
%primarily focus on $\alpha-\Omega$ dynamos because many astrophysical
%objects typically have significant differential rotation
% (Moffatt 1978; Parker 1979; Krause \& R\"adler 1989).
%But these textbook dynamos
 do not conserve magnetic helicity and this has rendered  them unable to predict how  dynamos saturate.  Their deficiency in this regard is conspicuous even in their diagrammatic representations
when the large scale magnetic field as represented as a 1-D line.
The basic notion that ribbons better represent the magnetic field for large scale dynamos 
has been emphasized before (Blackman \& Brandenburg 2003), but here we 
 incorporate this principle more  specifically  into a  full description and schematic of the $\alpha^2$ dynamo.
 The absence of differential rotation makes the $\alpha^2$  dynamo even simpler than the common $\alpha-\Omega$ dynamo, but
 it is still very much sufficient to illustrate the essential principles.

\begin{figure}
\minipage{.5\textwidth}
  \includegraphics[width=\linewidth]{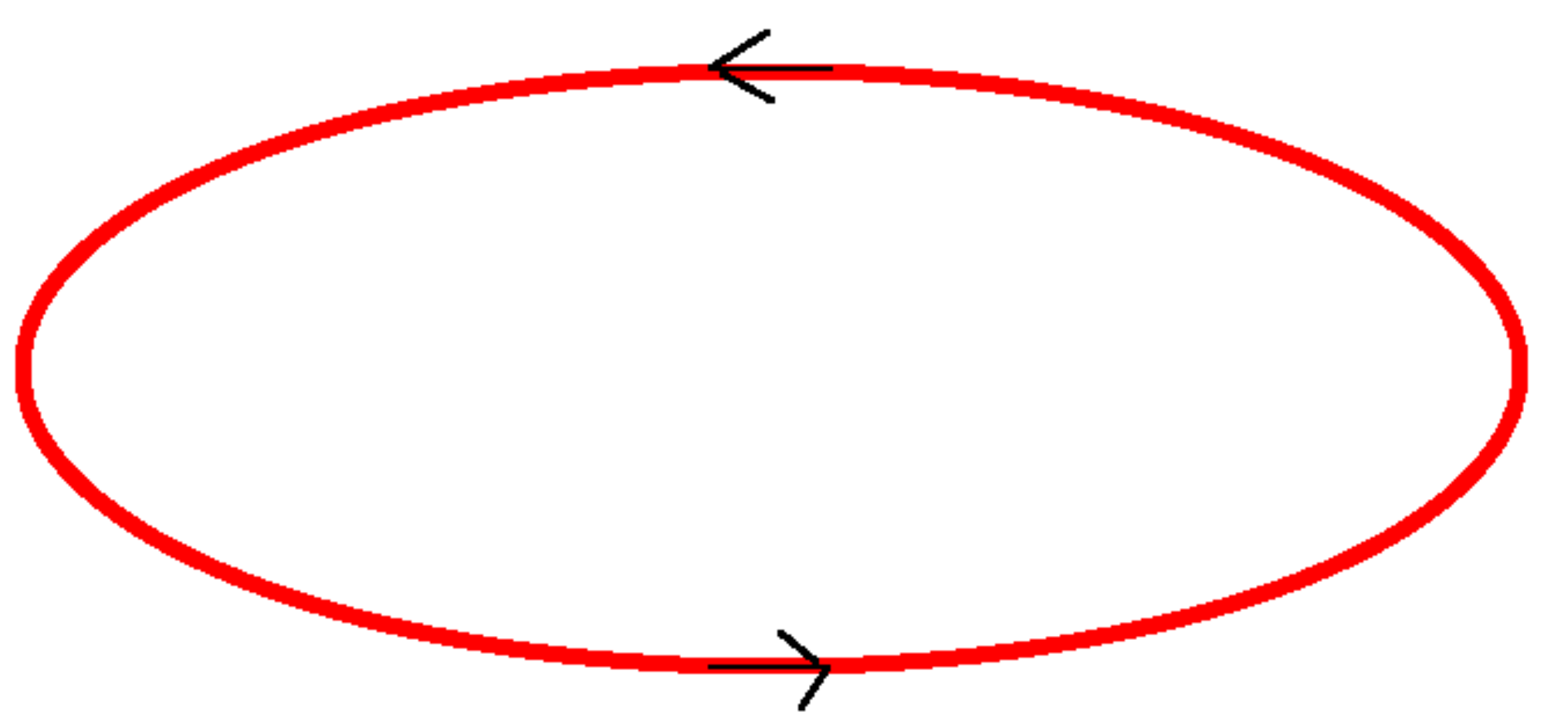}
 % \caption{A really Awesome Image}\label{fig:awesome_image1}
\endminipage\hfill
\minipage{.6\textwidth}
  \includegraphics[width=\linewidth]{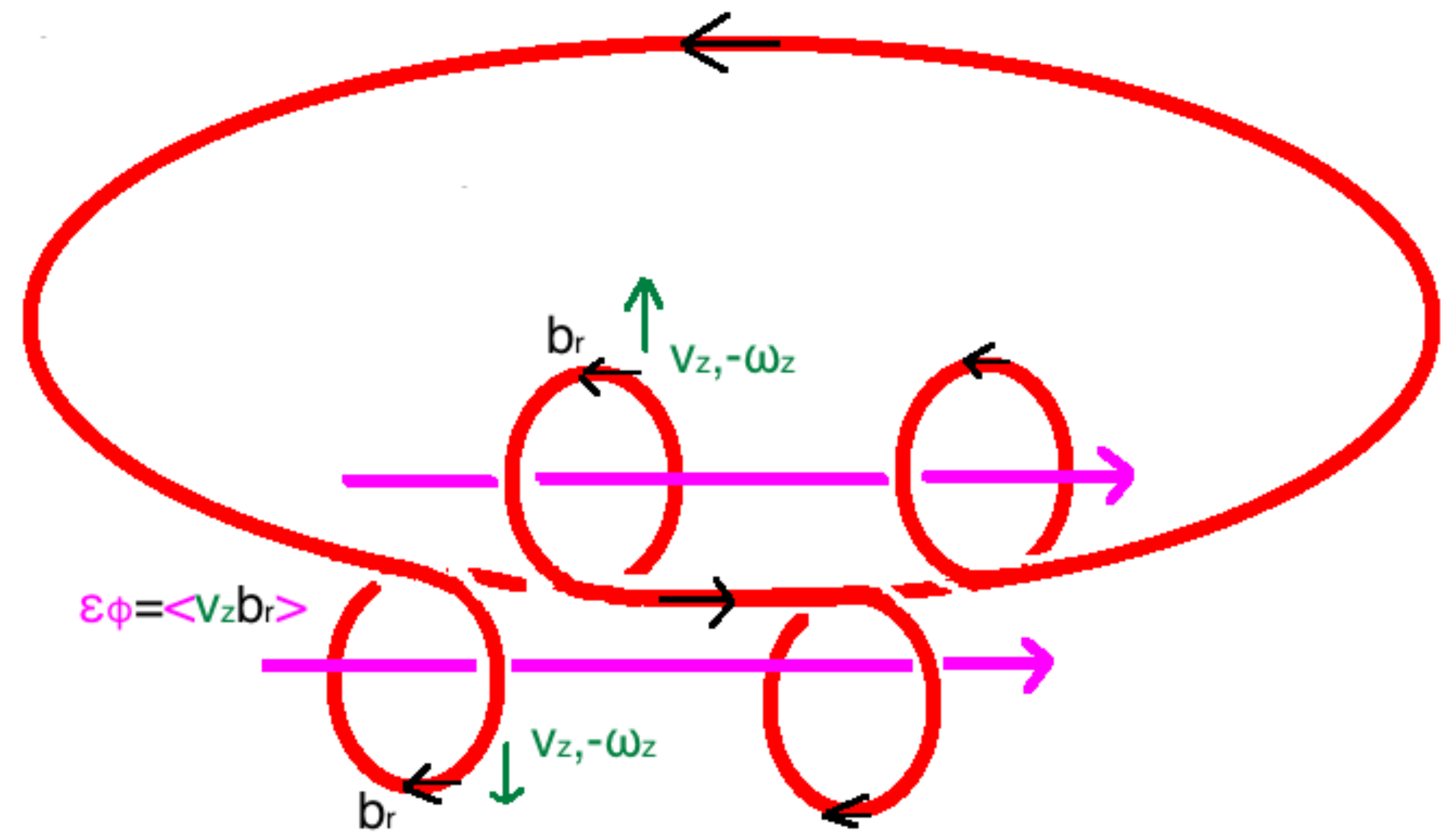}
%  \caption{A really Awesome Image}\label{fig:awesome_image2}
\endminipage\hfill
\minipage{.7\textwidth}%
  \includegraphics[width=\linewidth]{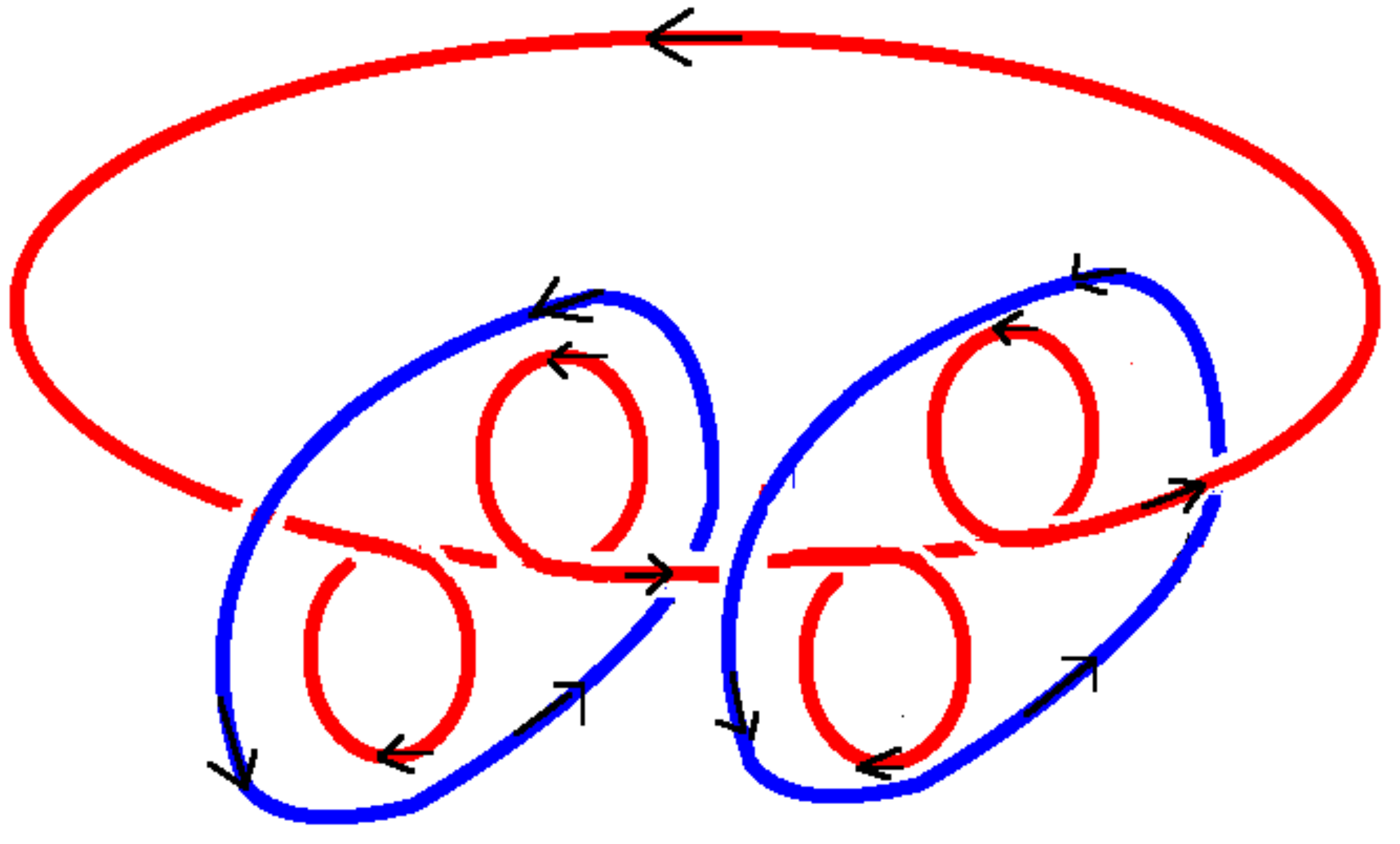}
%  \caption{A really Awesome Image}\label{fig:awesome_image3}
\endminipage
\caption{Two-stage schematic for  magnetic field structure in an  $\alpha^2$ dynamo driven by negative  kinetic helicity in the conventional 20th century approach with the magnetic field represented as lines.
  The first   panel shows a large untwisted  toroidal magnetic field ribbon.
The second panel shows the action of  kinetic helicity on the initial ribbon. The small scale negative kinetic helicity produces each of the four small scale poloidal loops.  Each small loop  incurs a writhe (or overlap) of positive (=right-handed) magnetic helicity.   The direction of the mean electromotive force is shown in the second panel. It is parallel to the mean poloidal current and highlights that both top and bottom loops have right handed writhe.
  In the third panel, the two intermediate scale poloidal loops  encircling the small scale loops represent the resultant mean poloidal field 
averaged   separately over each pair of  loops.  These intermediate scale  loops are  linked to the initial large scale torioidal loop.  Since a single linked pair of ribbons has  2 units of  magnetic helicity we see that the two poloidal loops linking the torioidal loop have a total 4 total units of right handed magnetic helicity. This helicity in the ``large scale field'' has come from zero initial magnetic helicity
and thus cannot be correct for MHD at large $R_M$ which conserves magnetic helicity. Therefore this diagram does not 
not account for the missing small scale magnetic helicity of opposite sign. (Compare to Fig. 2)}
%EB3 edited caption above for new figure
\label{fig1}
\end{figure}

Fig. \ref{fig1}  shows the amplification of large scale magnetic fields by the $\alpha^2$ dynamo  based on the traditional representation of the field as a 1-D line.  From a single initial toroidal  loop,  four small scale helical eddies, each with typical velocity $\bfv$ and kinetic helicity $\bfv\cdot (\curl \bfv) < 0$, create four small scale poloidal loops  with right handed writhe.  The result is a  toroidal EMF  $\emf_\phi =\lb v_z b_r \rb $ of the same sign inside all of the loops, regardless of whether they are above or below the 
initial toroidal loop.   The vectors in the small scale loops are shown explicitly just for two loops to avoid clutter, 
and the resultant EMF vector is shown.
%EB3 tweaked above for new figure
The result is  two net large scale poloidal magnetic field loops (shown in blue) in the third panel.  
But since the magnetic helicity is  a measure of linkage,  there is a problem:
The first panel has only   a single red loop, but seemingly evolves to a configuration with the red loop linked to the
two blue loops.  The system has somehow transformed from having no helicity to having  two units of linkage helicity. This  non-conservation
of helicity is inconsistent with the equations of MHD.

\begin{figure}
\minipage{.5\textwidth}
  \includegraphics[width=\linewidth]{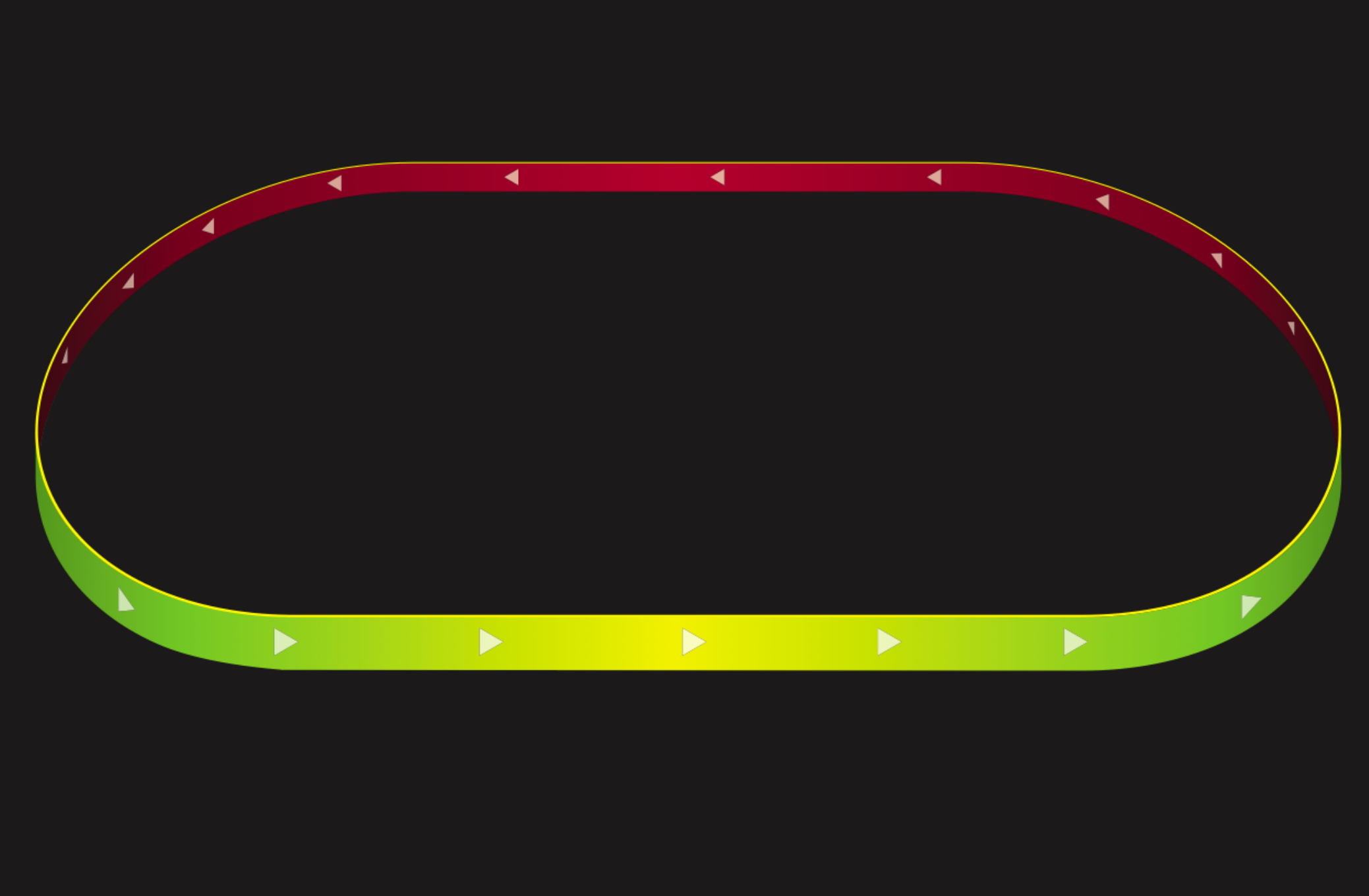}
 % \caption{A really Awesome Image}\label{fig:awesome_image1}
\endminipage\hfill
\minipage{.52\textwidth}
  \includegraphics[width=\linewidth]{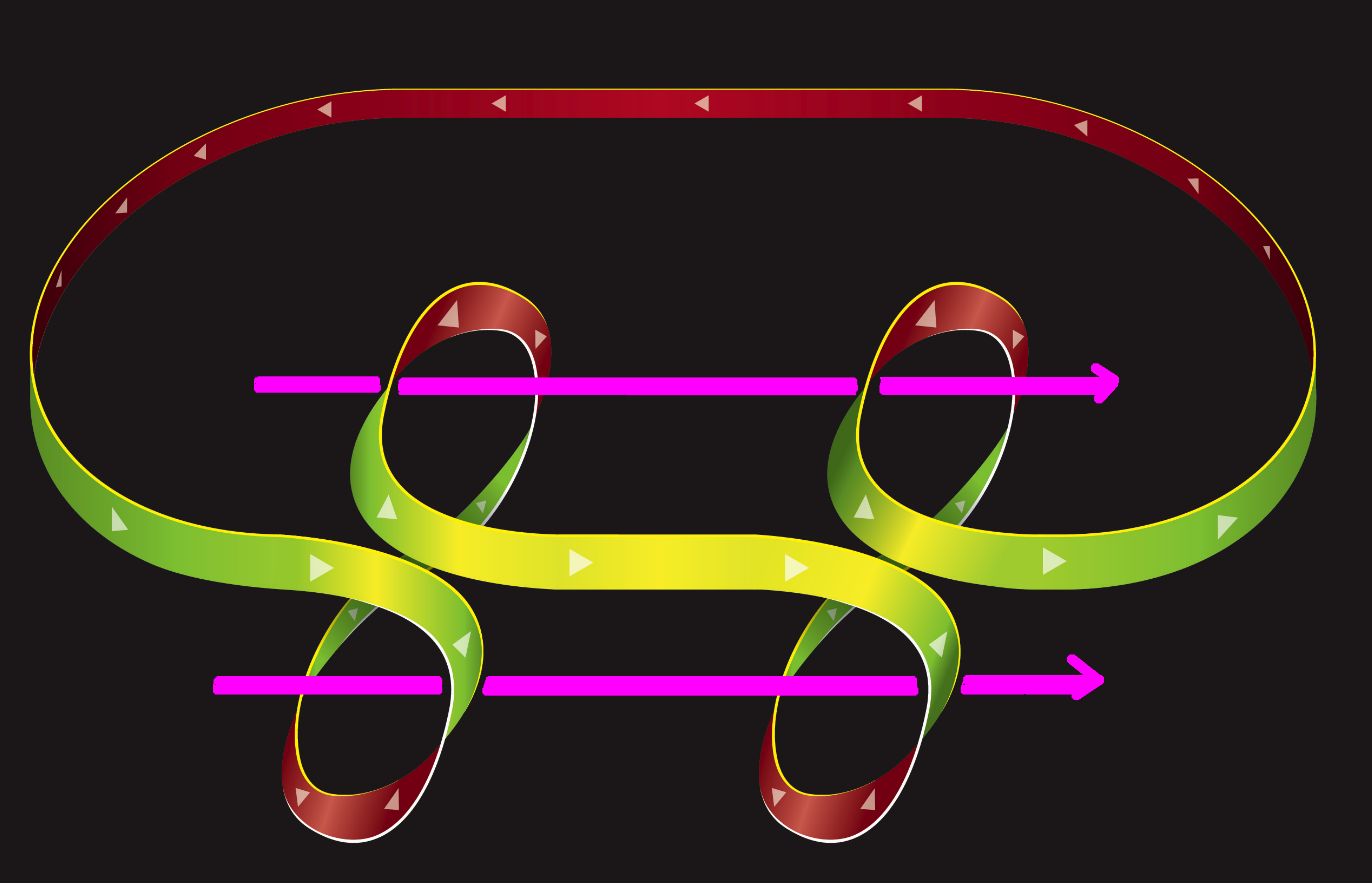}
%  \caption{A really Awesome Image}\label{fig:awesome_image2}
\endminipage\hfill
\minipage{.8\textwidth}%
    \includegraphics[width=\linewidth]{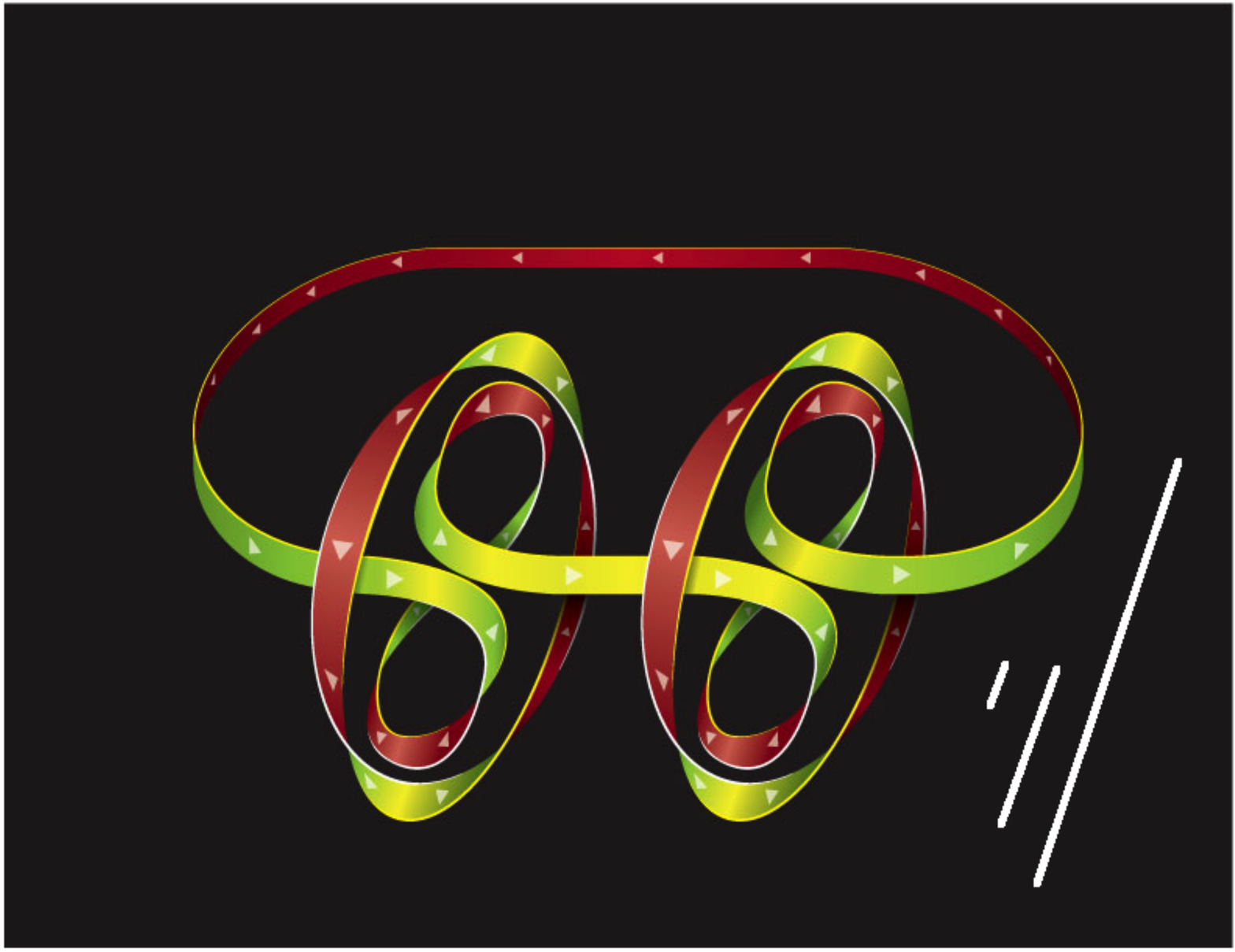}
%  \caption{A really Awesome Image}\label{fig:awesome_image3}
\endminipage
\caption{Correction of Fig 1 to include magnetic helicity conservation. The top panel shows a large untwisted  toroidal magnetic field ribbon.
The bottom shows the action of  kinetic helicity on the initial ribbon. The small scale negative kinetic helicity produces each  the four small scale poloidal loops.  The triangular arrows indicate magnetic field direction.
Each small loop   incurs a writhe (or overlap) of positive (=right-handed) magnetic helicity. Since magnetic helicity is conserved, each of these four loops also has a negative (=left-handed) twist along the field ribbon.   The two intermediate scale poloidal loops  encircling the small scale loops represent the resultant mean poloidal field 
averaged   separately over each pair of  loops.  
The intermediate scale poloidal loops have  accumulated two units of magnetic twist, one from each of the small loops that they encircle. These intermediate scale  loops are also  linked to the initial large scale toroidal loop.  Since a single linked pair of ribbons has  2 units of  magnetic helicity we see that the two poloidal loops linking the toroidal loop have a total 4 total units of right handed magnetic helicity in linkage which exactly balances the sum of 2+2 left handed units of twists on these poloidal loops.  In general, the small scale of the twists need not correspond to the same scale as the velocity driving the small scale writhes, though the calculations of section 4 assume such for simplicity. As in Fig 1. the  poles threading the  loops in the second panel are  parallel to the EMF and the mean poloidal current,  and serve as a visual tool to clarify that both top and bottom loops have right handed writhe.  
%{\bf 
The three white lines in the bottom panel indicate the approximate relevant length scales from smallest to largest $k_{2}^{-1}$ (ribbon width), $k_{f}^{-1}$ (inner loop size), and $k_{1}^{-1}$ (outer loop size), respectively, which appear in 
Eqs.   (\ref{9}-\ref{11}). }
%}
%EB3 edited caption above for new figure
\label{fig2}
\end{figure}

The solution to this conundrum  is shown in Fig. \ref{fig2}.   The analogous  diagrams as in Fig. \ref{fig1} are shown with the field 
represented by ribbons instead of lines.
Conservation of helicity is now  maintained:  the second panel shows how the right handed writhe of the small poloidal loops is compensated  by the opposite (left handed) twist helicity along the loops. The linkage that results in the third panel is in turn compensated   by the exact opposite amount of small scale twist helicity along the loops.  
Specifically the intermediate scale poloidal loops have  acquired two units of magnetic twist, one from each of the small loops that they encircle. These intermediate scale  loops are also  linked to the initial large scale toroidal loop.  Since a single linked pair of ribbons has  2 units of  magnetic helicity, we see that the two poloidal loops linking the toroidal loop have a total 4 total units of right handed linkage magnetic helicity  which exactly balances the sum of 2+2 left handed units of twist on these poloidal loops. 
%{\bf 
The comparison of Figs. 1 and 2 is also harmonious with the comparison of Figs. 8 and 9 of Pfister \& Gekelman (1991)
showing the non-conservation of magnetic helicity of reconnecting loops when they are treated as lines, and the
conservation when they are treated as ribbons.
%}

Visualizing the field as a 2-D ribbon rather than a 1-D line also illustrates the conceptual key to understanding  nonlinear quenching of the $\alpha^2$ dynamo.  In a system driven with kinetic helicity, the back reaction on the driving flow comes  from the buildup of the small scale twist. Field lines with small scale twists become harder to bend because the small scale current $\bfj$ associated with the twist  leads to  a  Lorentz force   $\bfj\times \bbB$, that pushes back against the small scale driving flow. 
Only the helical part of the small scale field enters the back reaction for flows which have at most weak anisotropy.  
This fact is  consistent with the equations discussed in the  following section, the original  spectral approach of Pouquet et al. (1976), and numerical simulations (Brandenburg 2001; Park \& Blackman 2012).
%{\bf
 As discussed further in Sec. 5, once the small scale helicity builds up significantly, the $\alpha^2$ dynamo growth rate slows
to the point where the near exact conservation of magnetic helicity is violated and resistive terms become important.
But it is the helicity conserving phase which causes the back reaction that leads to the transition.
%}

  Much recent work on astrophysical dynamos has been focused on how to sustain the EMF such that the  quenching from a build up of small scale helicity  is avoided and the EMF  is instead sustained by global or local helicity fluxes (the distinction is reviewed in Blackman 2014). In Fig 2, global helicity fluxes would mean removing the constraint that the ribbon is closed, and opening the boundaries over the region where the EMF is calculated  to allow  fluxes to remove the small scale twists.   Blackman \& Brandenburg (2003) showed a schematic of how this might work in the context of Coronal Mass Ejections of the sun, with the flux ejection both being important for sustaining  a fast dynamo, and also providing energy for particle acceleration in the corona.   Opening up the boundaries to allow the small scale twist (as opposed to large scale writhe) to preferentially leak away allows larger saturated field strengths and a longer regime of $R_M$ independent growth for the $\alpha^2$ dynamo (Blackman 2003).
For the $\alpha-\Omega$ dynamo that include magnetic  helicity dynamics, the role of global or local  helicity fluxes   is particularly  germane because the large scale field  can decay catastrophically in their absence in some models (e.g. Shukurov et al. 2006, Sur et al. 2007) after an initial transient growth phase. See also Hubbard \& Brandenburg (2012) for a discussion of possibly less catastrophic decay even without global fluxes, and Ebramhi \& Bhattacherjee (2014) for direct numerical evidence of local fluxes sustaining a large scale dynamo in sheared system.
%AH: ?
%EB fixed ok?
%AH: I'd at least like a counter-cite to Hubbard&Brandenburg 2012
%EB3: indeed...see the revision above.
The role of helicity fluxes as sustainers of the EMF as a matter of principle (e.g. Blackman \& Field 2000a) and specifically for sheared rotators 
 (Vishniac \& Cho 2001; Brandenburg \& Subramanian 2005a; Shukurov et al. 2006; Vishniac 2009; K\"apyl\"a \& Korpi 2011 Ebhami \& Bhattacharjee 2014;Vishniac \& Shapovalov 2014) is a topic of  active research.
Note that the role of helicity fluxes in sustaining  laboratory plasma dynamos in fusion devices has long been studied (e.g. Strauss 1985; 1986; Bhattacharhee \& Hameri 1986).

Fig. \ref{fig2}  also highlights  that the  scale of the twists along the ribbon can be different from the scale of the turbulent forcing, the latter of which may better represented by  the radius  of the writhed loop.
The reason for such scale separation in a real system would likely depend on the particular circumstance, such as  the ratio of vertical to radial density gradient scales.  However to give a sense of how such scale separation might arise even for an isotropic (but reflection asymmetric) system 
recall  that the diagram  is intended  to capture helical forcing with
$\langle {\bf v} \cdot { \bfw}\rangle < 0, $
where
$\bfw \equiv \nabla \times \bfv$
is the eddy vorticity.   
The  velocity gradient scale, and thus the scale of kinetic helicity would be that of the eddy itself.  
But depending on the vorticity profile within the  eddy, magnetic  flux at near the eddy core can be dissipated via the large field gradients there and preferentially amplified  by shear toward  the eddy's edge (Weiss 1966). 
%EB2 tweaked above to address your comment about expulsion
Neighboring eddies would have a similar effect, and could overall drive the system into a kind of sponge with  the magnetic field filling fraction potentially as small as the ratio of the average magnetic pressure to the average thermal pressure (Blackman 1996).   Once a vortex has partly expulsed magnetic flux  into a thinner structure (i.e. ribbon),  two scales in the magnetic structure now emerge as a result of the additional component of velocity parallel to the vorticity:  First, the writhe of the ribbon would be determined by the overall size of the eddy (the ``hole'' where there is less magnetic flux). Second,  this  velocity component   will ``roll''  the ribbon as it writhes,  twisting it with helicity sign opposite to the writhe. Since the ribbon would be thin compared to the eddy scale, the resulting twist would be of a  smaller scale than the
overall eddy (forcing) and writhe scales.

\section{ Equations of The $\alpha^2$ Dynamo}

To make the connection between the diagram of the $\alpha^2$ dynamo of Fig. 2  and the  dynamo equations,   we   follow  standard  derivations of the mean field equations for the $\alpha^2$ dynamo that incorporate magnetic helicity conservation
%(most closely  \citet{2013MNRAS.429.1398B})
%AH: meaning preserved?
%EB3: good
(closely following \citet{2013MNRAS.429.1398B}),
 but  deviate where noted below, to allow a scale separation between the forcing scale and the small scale current helicity buildup scale.

The electric field is
\beq
\bfE=-\nabla\Phi -{1\over c}
\partial_t\bfA,
\label{1b}
\ee
where $\Phi$ is the
%and $\bfA$ are the vector potentials.
scalar potential. Taking an average (spatial, temporal, or ensemble), and denoting averaged
values by the overbar, we have  
\beq
\bbE=-\nabla{\overline \Phi} -{1\over c}\partial_t\bbA
\label{2b}
\ee
Subtracting (\ref{2b}) from (\ref{1b}) gives the equation
for the fluctuating electric field
\beq
\bfe=-\nabla\phi -{1\over c}\partial_t\bfa
\label{3b},
\ee
where $\phi$ and $\bfa$ are the fluctuating scalar and vector potentials.
Using 
$\bbB\cdot \partial_t \bbA= \partial_t(\bbA\cdot \bbB) +c\bbE\cdot \bbB -c\nabla \cdot (\bbA\ts \bbE)$,
where the latter two terms result 
from Maxwell's equation $\partial_t \bbB=-c\curl \bbE$,
and the identity 
$\bbA \cdot \curl \bbE = \bbE\cdot\bbB-\nabla \cdot (\bbA \ts \bbE)$, 
we take the dot product of (\ref{2b}) with $\bfB$  and  obtain
\beq
\partial_t(\bbA\cdot\bbB)= -2c\bbE\cdot\bbB
%-\div ({\overline\Phi}\ \bbB + c\bbE\ts \bbA).
-\div ({c\overline\Phi}\ \bbB + c\bbE\ts \bbA).
\label{5b}
\ee
Similarly, by dotting (\ref{3b}) 
with  $\bfb$, the 
evolution of the
mean helicity density associated with  fluctuating fields is
\beq
\partial_t\overline{\bfa\cdot\bfb}= -2c\overline{\bfe\cdot\bfb}
%-\div(\overline{{\phi} \bfb} + c\overline{\bfe\ts \bfa}).
-\div(c\overline{{\phi} \bfb} + c\overline{\bfe\ts \bfa}).
\label{6b}
\ee

To  eliminate the electric fields 
from (\ref{5b}) and (\ref{6b}) we use Ohm's law with  a resistive term to obtain
 \beq
{\bfE}=-\bfV\ts\bfB/c +\eta \bfJ,
\label{7b}
\ee
where $\bfJ={c\over 4\pi}\curl \bfB$ is the current density and $\eta $ is the resistivity. 
 Taking the average gives 
\beq
{\bbE}=-\emfb/c -\bbV\ts\bbB/c+\eta \bbJ,
\label{8b}
\ee
where $\emfb\equiv \overline{\bfv\ts\bfb}$ is the turbulent electromotive
force.
Subtracting (\ref{8b}) from (\ref{7b}) gives  
\beq
{\bfe}=(\emfb-\bfv\ts\bfb -\bfv\ts\bbB -\bbV\ts\bfb)/c 
+\eta \bfj.
\label{9b}
\ee

Plugging  (\ref{8b}) into (\ref{5b}) and (\ref{9b}) into (\ref{6b}) and  now globally averaging (indicated by brackets)  
and assuming divergence terms  then vanish then gives
\beq
{1\over 2}\partial_t\lb\overline{\bfa\cdot\bfb}\rb=
-\lb\emfb\cdot\bbB\rb
-\nu_M \lb\overline{ \bfb\cdot\curl\bfb}\rb,
\label{5aa}
\ee
%where $\nu_M= {\eta c^2\over 4\pi }$.
where $\nu_M\equiv (\eta c^2/4\pi)$.
and
\beq
{1\over 2}\partial_t\lb\bbA\cdot\bbB\rb=\lb\emfb\cdot\bbB\rb-\nu_M\lb\bbB\cdot\curl\bbB\rb.
\label{6aa}
\eeq
%AH: add
%EB3: i like it. i put in parenthesis though below and moved the triply periodic box comment therein
(Note that such globally averaging only eliminates divergence terms if the system goes to zero at infinity or otherwise is appropriately symmetric, such as a  triply periodic simulation box.  A shearing sheet's shear velocity goes to infinity at $\pm \infty$, so $\alpha - \Omega$ dynamos
in shearing sheets cannot be so simply treated.)

To obtain an expression for $\emfb$, we use the  `minimal tau' 
closure  for incompressible MHD \citep{2002PhRvL..89z5007B, 2005PhR...417....1B} to replace  triple correlations by a damping term on the grounds that  $\emfb$ should decay in the absence of $\bB$.  For a system which can be reflection asymmetric but at most only weakly anisotropic,  
this gives 
\beq
\partial_t \emfb = \lb \partial_t \bfv  \times \bfb \rb + \lb \bfv \times \partial_t \bfb \rb 
={\alpha \over {\tau}}\bbB-{\beta\over {\tau}}\curl\bbB-\emfb/{\tau},
\label{7aa}
\ee
where ${\tau}$ is a damping time 
and 
\[
\alpha \equiv {{\tau}\over 3} \left({\lb \bfb\cdot\curl\bfb \rb\over 4\pi\rho}- 
\lb \bfv\cdot\curl\bfv \rb\right) \quad  {\rm and} \quad 
\beta\equiv   {{\tau} \over 3}\lb v^2\rb.
\]
%AH: add
The $\alpha$ above is the $\alpha$ effect that drives $\alpha^2$ and $\alpha - \Omega$ dynamos, while the
$\beta$ above is the turbulent diffusivity.
%EB3: i changed resisitivity to diffusivity above

The time evolution of  $\emfb$ can be retained as a separate equation, but  simulations of magnetic field evolution in
forced isotropic helical turbulence have shown both that the simulations are well matched when the left side of (\ref{7aa}) is ignored  and 
that ${\tau}\sim {1\over v_f k_f}$, the eddy turnover time associated with the forcing scale
 \citep{2002ApJ...572..685F,2005A&A...439..835B}. 
Rearranging  (\ref{7aa}) with this approximation then gives
\beq
\emfb =\alpha \bbB-\beta \curl\bbB,
\label{emfb}
\ee
Eqs. (\ref{5b}) and (\ref{6b}) then become
\beq
{1\over 2}\partial_t\lb\overline{\bfa\cdot\bfb}\rb=
%-\alpha\bB^2 +\beta\bbB\cdot\curl\bbB
-\alpha\lb\bB^2\rb +\beta\lb\bbB\cdot\curl\bbB\rb
-\nu_M \lb\overline{ \bfb\cdot\curl\bfb}\rb
\label{5ab}
\ee
and
\beq
{1\over 2}\partial_t\lb\bbA\cdot\bbB\rb=
%\alpha\bB^2 -\beta\bbB\cdot\curl\bbB
\alpha\lb\bB^2\rb -\beta\lb\bbB\cdot\curl\bbB\rb
- \nu_M\lb\bbB\cdot\curl\bbB\rb.
\label{6ab}
\ee
The  energy associated with the small scale magnetic field does not enter  $\emfb$  so 
it does not  enter equations (\ref{5aa}) and (\ref{6aa}). It enters as a higher order  correction \citep{2003PhRvL..90x5003S}  which  we presently ignore.
 %because its ratio to the $\beta$ term in the EMF is  ${ b^2\over 4\pi \rho v^2}{k_1^2\over k_2^2}<<1$.  
However, upon plugging (\ref{emfb}) into (\ref{5aa}) and (\ref{6aa}) , the energy associated with the large scale field $\bB^2$ {\it does} enter.
Therefore we need a separate  equation
for the energy associated with the energy of the mean field.
To obtain this equation we dot $\partial_t \bbB = -c\curl \bbE$
with $\bbB$ and ignore the flux terms to obtain
\beq 
\begin{array}{r}
{1\over 2}\partial_t \lb \bB^2\rb =-c\lb \bbB\cdot \curl \bbE\rb=
-c\lb \bbE\cdot \curl\bbB\rb
=\lb\emfb\cdot \curl\bbB \rb
-\nu_M\lb (\curl\bbB)^2 \rb
=
\alpha \lb \bbB\cdot\curl \bbB\rb
 - \beta \lb (\curl\bbB)^2 \rb
-\nu_M\lb (\curl\bbB)^2 \rb,
\end{array}
\label{ind}
\ee
where the latter two similarities follow from using
(\ref{8b}) and (\ref{emfb}) and 
$\bbV=0$.

% The energy associated with the large scale field $\bB^2$ {\it does} enter the last two equations and we would in general need a separate equation for it.
%However, in the case when kinetic forcing is maximally helical, the large scale
%magnetic field grows to be almost fully  helical for most (but not all) of the evolution of the $\alpha^2$ dynamo  to saturation (Park and Blackman 2012).
%Thus we assume 
%\beq \langle \overline B ^2\rangle /|\langle\bbA\cdot\bbB\rangle| = {\rm  constant}
%\label{fulhel}
%\eeq
%Then, in the discrete scale approach below, 
%the  large scale magnetic energy equation is redundant with the large scale helicity equation.   

Eqs.  (\ref{5ab}), (\ref{6ab}), and  (\ref{ind})  form a set of equations that can be solved in a two scale model.
To facilitate study of the essential implications of this  coupled system   for a closed or periodic system, we adopt a discrete scale
approximation
  \citep{2002PhRvL..89z5007B,2005PhR...417....1B} 
 and indicate  large scale mean magnetic  
 quantities with  subscript ``1'', small scale magnetic quantities with ``2'', and the kinetic  forcing scale with the subscript $f$. 
We assume that  the  wave number  $k_1$    associated with the spatial variation scale of large scale quantities  satisfies  $k_1<< k_2$
and $k_1<<k_f$.   In the standard two-scale model for the $\alpha^2$ dynamo,  the kinetic forcing wavenumber $k_f=k_2$.
Here we allow $k_f$ and $k_2$ to be distinct, a small generalization motivated by our schematic diagrams
that provides some conceptual versatility.

Applying these scaling  approximations  to the  closed or periodic system,  we   then  use
 $\lb \bbB \cdot \curl \bbB \rb = k_1^2 \lb \bbA \cdot \bbB \rb\equiv k_1^2 H_1$ where $H_1$ is the magnetic helicity
 of the large scale magnetic field,  $\lb \bbB^2 \rb = k_1 |H_1|$,
         along with   
   $\lb \bfb\cdot \curl \bfb \rb = k_2^2 \lb \bfa \cdot \bfb \rb=k_2^2 H_2$ where $H_2$ is the magnetic helicity
   of the small scale field, and we assume that $H_1> 0$ for the cases discussed.
   %AH: these need a reference at least.  
   %EB3: i added again just the BS 2013 as its closest in form to  that because of the B_1 equation (though it goes back to bf2002 and such)
   Then  Eqs (\ref{5ab}) and (\ref{6ab}) become  (Blackman \& Subramanian 2013)
\beq
%\partial_t H_1= \left({2{\tau}\over 3}\right)k_2^2 H_2{B_1^2\over 4\pi \rho} - {2{\tau}\over 3} v_2^2k_1^2 H_1-2\nu_M k_1^2 H_1,
\partial_t H_1= \left({2{\tau}\over 3}\right)\left( {k_2^2H_2\over 4\pi \rho}-H_V\right){B_1^2} - {2{\tau}\over 3} v_f^2k_1^2 H_1-2\nu_M k_1^2 H_1,
\label{6}
\eeq
\beq
\partial_t H_2= -\left({2{\tau}\over 3}\right)\left({k_2^2 H_2\over 4\pi \rho}-H_V\right){B_1^2} + {2{\tau}\over 3} v_f^2k_1^2 H_1-2\nu_M k_2^2 H_2,
\label{7}
\eeq
and
\beq
\partial_t B_1^2 = \left({2{ \tau}\over 3}\right)\left({k_2^2 H_2 \over 4\pi \rho} -H_V \right)k_1^2 H_1 - {2{ \tau}\over 3} v_f^2 k_1^2 B_1^2-2\nu_M k_1^2 B_1^2,
\label{8}
\eeq
where $H_V\equiv \lb \bfv\cdot\curl \bfv\rb\simeq |f_hk_f v_f^2|$  is the kinetic helicity driven into the system on the forcing scale, which for dimensionless $f_h=1$
would be maximally helical.

Note that since $\tau$ is the damping time from the minimal $\tau$ approximation (rather than an eddy turnover time),
$k_f$ does not yet explicitly enter the equations until we specific its value.
We now  non-dimensionalise these  equations by scaling lengths in units of $k_f^{-1}$
and time in units of $\tau$, where we assume $\tau=(k_f v_f)^{-1}$.
%AH: add, needs a citation:
This latter assumption is made on dimensional grounds, but has been  numerically validated  \citep{2005A&A...439..835B}.
%EB3 added above
We define
\[h_1\equiv {k_f H_1\over 4\pi \rho v_f^2}, \
h_2\equiv {k_f H_2\over 4\pi \rho v_f^2}, \
h_V\equiv {H_V\over  k_f v_f^2}, \
R_M\equiv {v_f\over \nu_M k_f}.
{\rm and}\  
M_1\equiv{B_1^2\over 4\pi \rho v_f^2}.
\]
Eqs. (\ref{6}),  (\ref{7}) and (\ref{8}) can then  be respectively written as
\beq
%\partial_\tau h_1 = {2\over 3}{k_1\over k_f}\left({k_2^2\over k_f^2}h_2-h_V\right) h_1-{2\over 3}\left({k_1\over k_f}\right)^2 h_1-{2\over R_M}\left({k_1\over k_f}\right)^2h_1,
\partial_\tau h_1 = {2\over 3}\left({k_2^2\over k_f^2}h_2-h_V\right) M_1-{2\over 3}\left({k_1\over k_f}\right)^2 h_1-{2\over R_M}\left({k_1\over k_f}\right)^2h_1,
\label{9}
\eeq
\beq
%\partial_\tau h_2 = -{2\over 3}{k_1\over k_f}\left({k_2^2\over k_f^2}h_2-h_V\right) h_1+{2\over 3}\left({k_1\over k_f}\right)^2 h_1-{2\over R_M}\left({k_2\over k_f}\right)^2h_2,
\partial_\tau h_2 = -{2\over 3}\left({k_2^2\over k_f^2}h_2-h_V\right)M_1+{2\over 3}\left({k_1\over k_f}\right)^2 h_1-{2\over R_M}\left({k_2\over k_f}\right)^2h_2,
\label{10}
\eeq
and 
\beq
\partial_\tau M_1 ={2 \over 3}  \left({k_2^2\over k_f^2}h_2-h_V\right)\left({k_1\over k_f}\right)^2h_1 - {2 \over 3}M_1 \left({k_1\over k_f}\right)^2
 - {2\over R_M} M_1\left({k_1\over k_f}\right)^2.
\label{11}
\eeq
%Thus we assume 

\section{Solutions}
\begin{figure}
\minipage{.5\textwidth}
  \includegraphics[width=\linewidth]{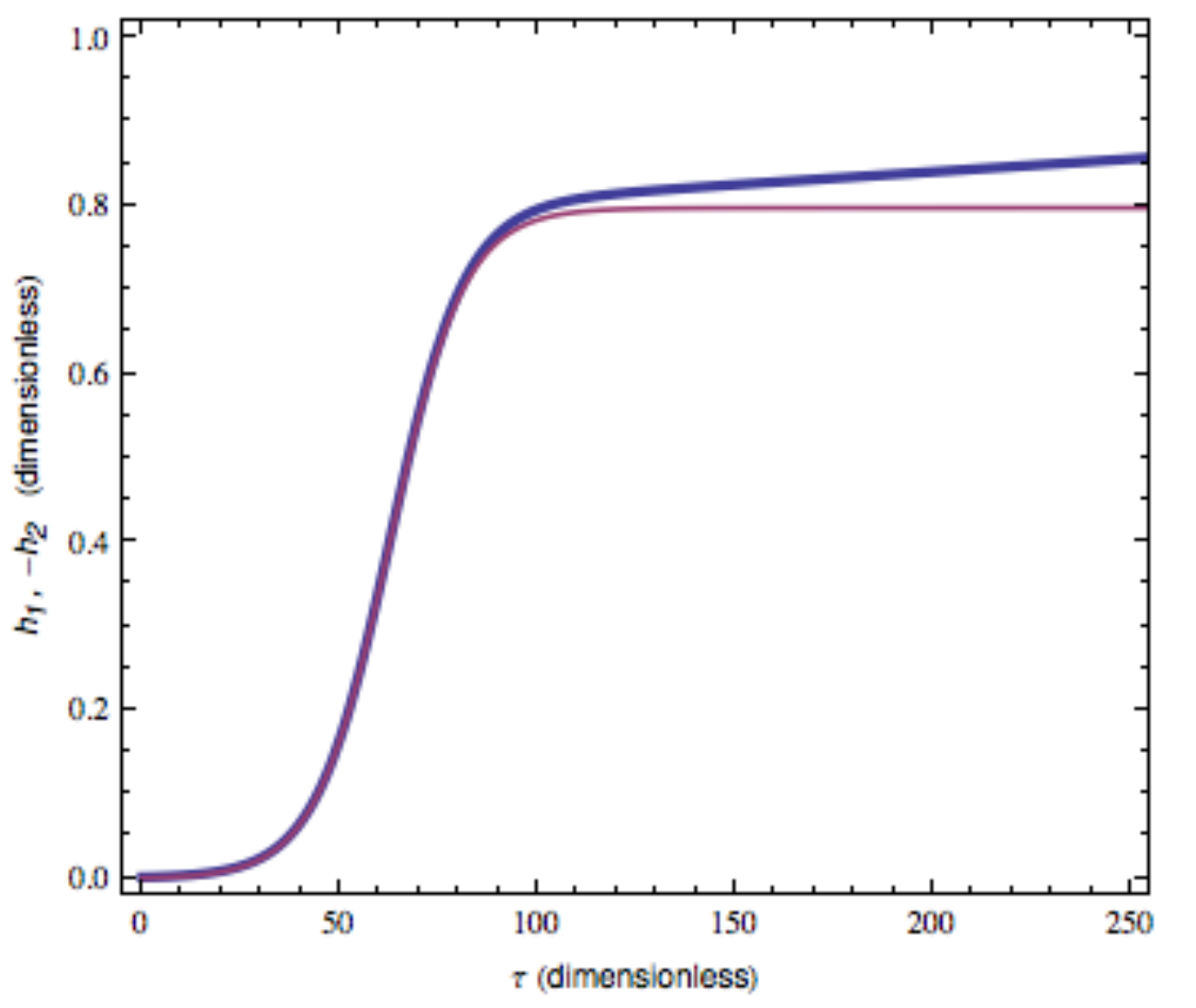}
 % \caption{A really Awesome Image}\label{fig:awesome_image1}
\endminipage\hfill
\minipage{.5\textwidth}
  \includegraphics[width=\linewidth]{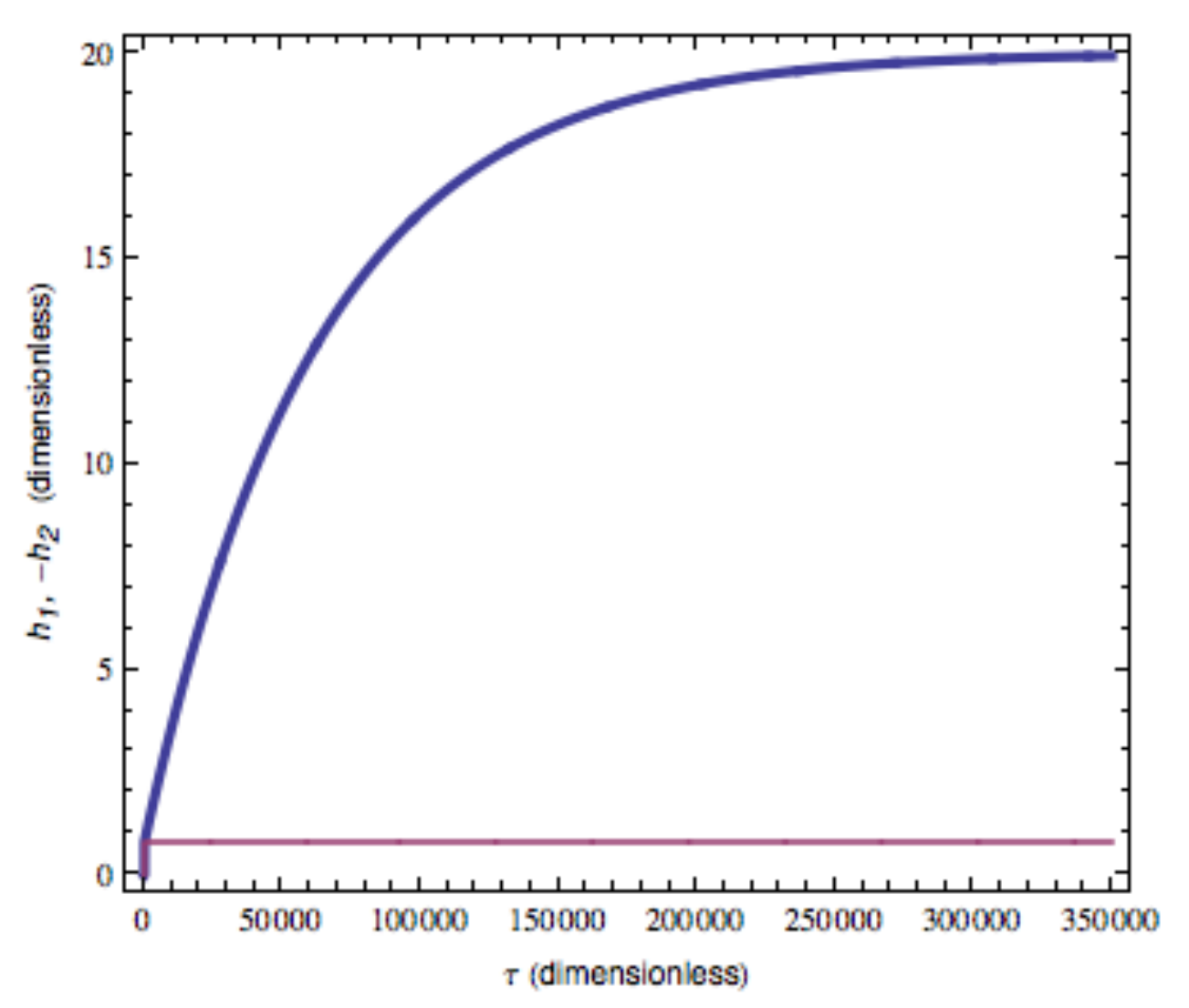}
%  \caption{A really Awesome Image}\label{fig:awesome_image2}
\endminipage\hfill
\minipage{.5\textwidth}%
  \includegraphics[width=\linewidth]{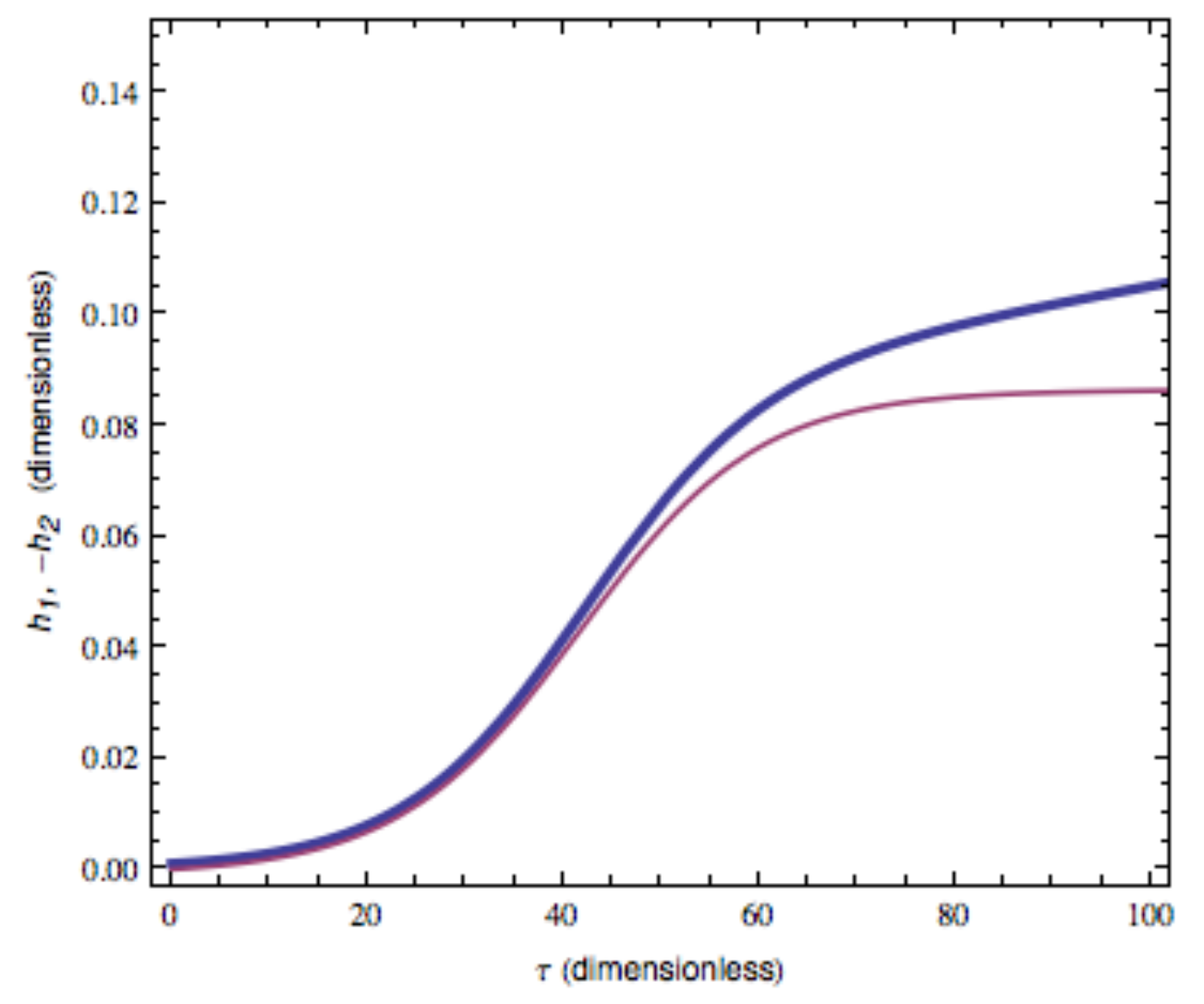}
  \endminipage\hfill
\minipage{.5\textwidth}%
  \includegraphics[width=\linewidth]{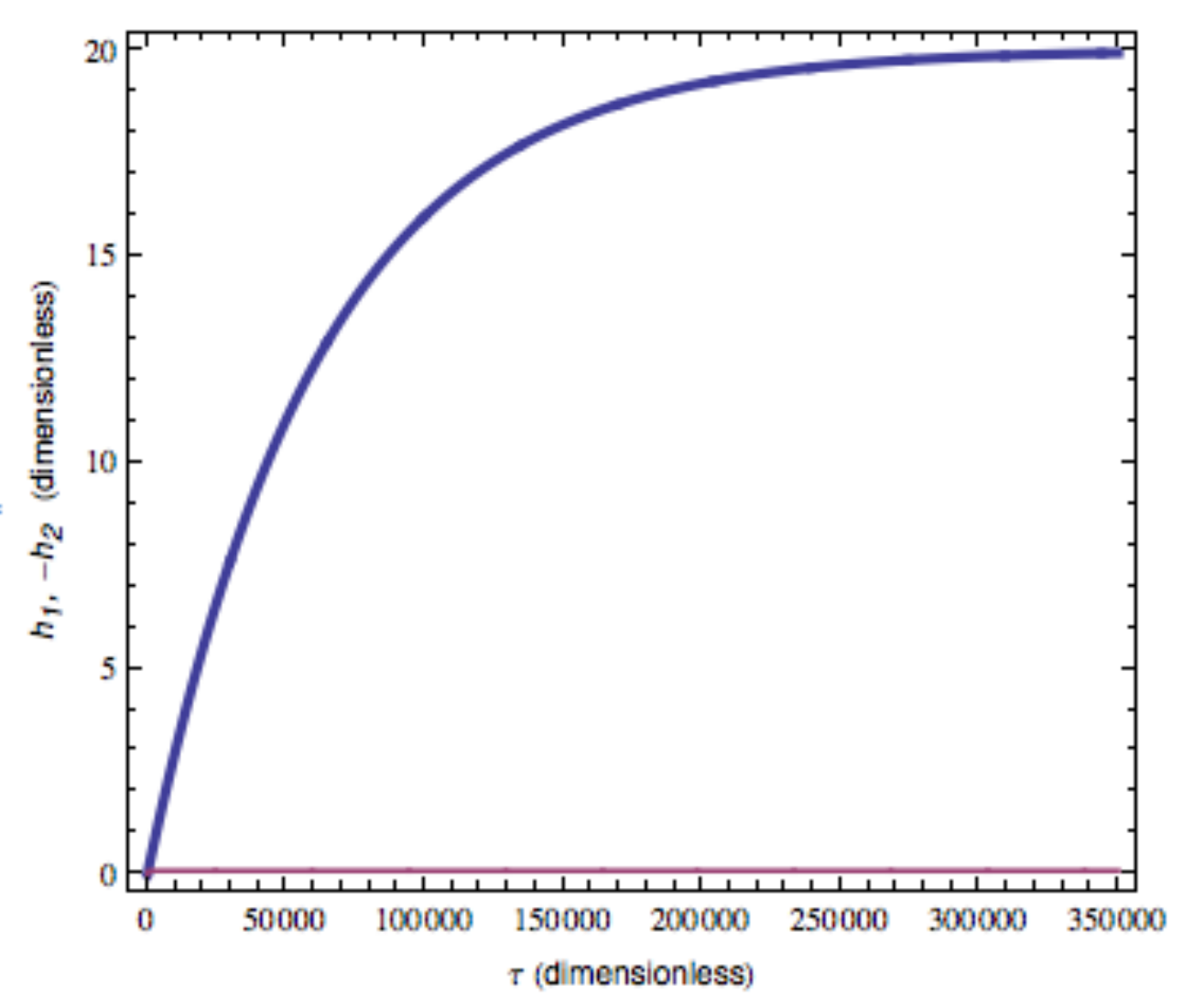}
 \endminipage
\caption{Solutions to Eqs. (\ref{9}), (\ref{10}), and  (\ref{11})  for the fully helical forcing case $h_v=-1$ for all times and initially $h_1(0)=0.001$ and  $h_2(0) =0$ 
with zero non-helical magnetic energy. In this  case  Eq. (\ref{11}) can be ignored as it is redundant with (\ref{9}). For all panels $k_f=5$ and
$R_M=5000$.  The top row  and the bottom row (c and d) differ in that   $k_2/k_f = 1$ for the top row and $k_2/k_f=3$ for the bottom row.
The left panel in each row is the early time solution of $h_1$ (thick line) and $-h_2$ (thin line)  where the growth is independent of  $R_M$
The right  panels show the late time evolution where the magnitude the small scale magnetic  helicity  $h_2$ has grown enough to offset the kinetic helicity driving sufficiently so that the   $R_M$ terms  become important.  The asymptotic  saturation value of $h_1$ is independent of $k_2$ but the transition value of $h_1$ when $R_M$ becomes important is reduced  by $(k_f/k_2)^2$ 
as is the saturation value of  $h_2$.  That $k_f$ and $k_2$ need not be equal is also captured by
in Fig 2. }
%EB removed fig 2b changed caption
\label{fig3}
\end{figure}

Equation (\ref{11})  can be ignored if the initial total magnetic energy is small and the forcing is sufficiently helical, for then only the helical large scale magnetic  energy grows significantly and the magnetic energy $M_1$  that appears in (\ref{9}), (\ref{10}) is that of the helical field, and can be replaced by
$M_1= {k_1\over k_f} h_1$. We employ this here and solve Eqs. (\ref{9}), (\ref{10}) for $k_f=5$, $k_1=1$ and $R_M=5000$
for cases with $k_2=k_f$ and $k_2=3k_f$ respectively.
The results  are shown in Fig. \ref{fig3}.
The top row shows solutions for $h_1$ and $h_2$ at early and late times respectively  when $k_2=k_f$.
The bottom row shows solutions for $h_1$(thick blue line)  and $h_2$ (thin purple lines)  at early and late times respectively  when $k_2/k_f=3$.  In comparing the two right panels, we see that although $h_1$ saturates at the same value, $h_2$ saturates
at a value reduced by a factor $(k_f/k_2)^2$ compared to the top.  This can be seen analytically from setting the left sides of Eq. (\ref{9}) and Eq. (\ref{10}) equal to zero for the steady-state solution, which shows that at saturation, $h_1=-(k_2/k_1)^2h_2$.
Plugging this relation back in  Eq. (\ref{9}) for the steady state and dropping the $R_M$ term  because it is small,  we then obtain 
$h_2=- {k_f^2\over k_2^2}\left(1-{k_1\over k_f}\right)= 20/k_2^2$  and
$h_1= {k_f^2\over k_1^2}\left(1-{k_1\over k_f}\right)$=20.

We can estimate the transition from the kinematic growth regime, to the $R_M$ dependent growth regime shown in the left panels
of Fig. \ref{fig3}  as follows:  Before the $R_M$ terms enter  in  Eq. (\ref{10}),  the magnetic helicity of the large and small scales must be essentially conserved, and so $h_1 \simeq -h_2$ until the end of this regime.  
%{\bf
 The  deviation from approximate conservation of magnetic helicity 
becomes substantial when the sum of the terms independent of $R_M$ deplete to such an extent  that  the $R_M$ terms become important. 
%}
For large $R_M$ we can thus estimate the transition  by just balancing the
$R_M$ independent terms on the right hand side seting $h_1=-h_2$. The result gives  $h_1\simeq - h_2 \simeq {k_f^2\over k_2^2}\left(1-k_1/k_f\right)=20/k_2^2$
This explains why the transition point for both $h_1$ and $h_2$ are reduced as a function of increasing $k_2/k_f$.

%{\bf
 The ultimate steady-state is sustained against resistive decay by the kinetic helicity  forcing. If the forcing is removed at some point during the steady-state regime, some of the  large scale magnetic helicity would annihilate  all of the 
small scale magnetic helicity, leaving a residual  magnetic helicity on the large scale which would then decay resistively.
The complementary cases of (i) driven magnetic relaxation, where magnetic helicity (rather than kinetic helicity)  is injected 
and relaxes to large scales, and (ii)  large scale helical field decay, which occurs only on resistive time scales for sufficient large initial
field strength are reviewed in Blackman (2014). See also Bhat et al. (2014) for simulations of case ii.
%}

The allowance for a scale distinction between $k_2$ and $k_f$ whilst still using just two dynamical equations
is an intermediate approach between those which  have restricted the solution of equations to strictly two scales, and those which have added a third time evolution equation for magnetic helicity on the smallest scale in addition that on the forcing scale (Blackman 2003).  In a sense this model can then be thought of as a 2.5 scale model.
As discussed in section 3, the diagrammatic representation of the $\alpha^2$ dynamo of Fig. 2 allows for this distinction as well as the  tracking of magnetic helicity because the field is treated as a 2-D ribbon.  
%{\bf 
There   is  evidence from 3-D numerical simulations that the small scale current  helicity in $\alpha^2$ dynamo simulations is not in fact peaked on the forcing scale but on smaller scales (Park \&  Blackman 2012).  Despite the fact that the full spectrum of scales
is present in simulations, the large scale,  the forcing scale and the current helicity weighted  wave number emerge as natural scales
from which to build a discrete scale mean field theory.
 There is  good opportunity for further numerical simulation studies of  the $\alpha^2$ dynamo to   assess the efficacy of equations (\ref{9}-\ref{11}) and also to dynamically predict the 
ratio of $k_2/k_1$ for a given choice of $k_f$. 
%}

A subtlety regarding the scale separation of $k_2$ and $k_f$  is that  the time constant $\tau$ entering Eqs. (\ref{6}) and (\ref{7}) is just the damping constant appearing in the last term of 
(\ref{7aa}), resulting from the "minimal tau" closure. 
This contrasts  the first order smoothing approximation (FOSA, e.g. Moffat 1978) where the corresponding $\tau$ in the $\alpha$ effect emerges only as a single approximation to the  two correlation times in the separate time integrals $\lb\int \bfv(t)\cdot\curl \bfv(t') dt'\rb$ and $\lb\int \bfj(t)\cdot\bfb(t')dt' \rb$
for the kinetic helicity and current helicity terms respectively. In FOSA, even if $\lb\bfj\cdot\bfb\rb$ is not peaked on the forcing scale, $\tau \lb\bfj\cdot\bfb\rb$ could  be if $\tau$ scales inversely with $k$,  for example as $k^{-2/3}$ in the Kolmogorov model.   FOSA could also lead to a circumstance in which the $\tau$'s multiplying the two terms of the $\alpha$ effect could be different. The present formalism of allowing for $k_2\ne k_f$ can also be alternatively  interpreted to   accommodate  this possibility.

%{\bf
 Finally, note that because Fig 2 shows exact helicity conservation, it is most rigorously applicable in the time regime of solution to 
Eqs. (\ref{9}-\ref{11}). before the $R_M$ terms become significant (and thus the regime in the panels of Fig. 3 before the curves diverge).
This is appropriate because it is the build up of the small scale helicity that in turn eventually saturates the dynamo and its because
of this nearly helicity conserving phase that the subsequent $R_M$ dependent phase emerges.  
%}

\section{Conclusions}

%{\bf
 We have shown how considering the magnetic field to be a 2-D ribbon rather than  1-D line  leads to a schematic diagram that correctly captures the conservation of magnetic helicity for the $\alpha^2$ dynamo during its evolution to saturation
 %}
 unlike traditional dynamo diagrams that treat the magnetic field as a line. 
 %This basic notion of replacing  field lines with ribbons to capture magnetic helicity  conservation in dynamos has been discussed before, but  here we have shown specifically how this works in  a complete  schematic of the $\alpha^2$ dynamo
In addition, the diagram  illustrates the need to allow for   a distinction of 3 scales:
   (i) the helical forcing scale  (ii)  the small scale magnetic helicity (ii) the large scale magnetic helicity.
   Toward this end, we introduced a simple generalization to  the  two-scale equations of the $\alpha^2$ dynamo to allow for the fact the the forcing  scale and and scale of  current helicity buildup may not be equal without having to introduce  a third dynamical equation.
  Solving these equations shows that when the  scale of the current helicity buildup is much smaller than that of the forcing scale (though still well above the resistive scale), the quenching of the dynamo is  exacerbated when compared to the case in which  these latter two scales are equal. This provides a simpler framework to study the separation of these scales than has been considered before (Park \& Blackman 2013).
  
In short, the near conservation of  magnetic helicity in MHD dynamos for a large $R_M$  closed system is well captured visually when magnetic  fields are represented by ribbons but poorly captured when the fields are represented as lines. The associated diagram of the $\alpha^2$ dynamo also captures the potential richness of an additional scale separation  which, as we have shown, affects predictions of dynamo saturation when incorporated into the theory.  {There is opportunity for future  work to assess the efficacy of the 2.5 scale model
of the $\alpha^2$ dynamo herein  with simulations and to develop a theory to predict $k_2/k_2$ for a given $k_f$.}

Because dynamos are so  representative of how magnetic fields and  flows interact in MHD, the lesson learned seemingly has very  broad implications for high $R_M$ MHD beyond that of dynamo theory:  {\it Magnetic fields are better  visualized as 2-D ribbons than 1-D lines. }
%{\bf
 Magnetic reconnection provides another example which bears this out \cite{1991AmJPh..59..497P}.
 %}

\section*{Acknowledgments} 
We thank  K. Subramanian  for related discussions.
EB acknowledges support from grant
  NSF grant  AST-1109285, and from the Simons Foundation.
    AH acknowledges support from  National Science Foundation, Cyberenabled Discovery Initiative grant AST08-35734, National Aeronautics and Space Administration grant NNX10AI42G (DSE), and a Kalbfleisch Fellowship from the American Museum of Natural History.    
  %EB3 edited above
  %AH: ask me to track down stuff for me.

%AH: it might help if the pole is extended through the right hand small scale loop.  More nitpicky, where the pole is in front
%AH: of the loop, it'd be nice to ``cut'' the loop as you've done for the other overlaps.
%EB3: see revised figures!

\bibliographystyle{mn2e}

\end{document}